\documentstyle[pra,multicol,aps]{revtex}

\begin{document}
\input{epsf.tex}
\epsfverbosetrue

\title{Coupled-mode theory for Bose-Einstein condensates}

\author{Elena A. Ostrovskaya$^1$, Yuri S. Kivshar$^{1}$, Mietek
Lisak$^{2}$,
Bjorn Hall$^{2}$, Federica Cattani$^{2}$, and Dan Anderson$^{2}$}
\address{${}^1$ Optical Sciences Centre, Australian National University,
Canberra ACT 0200, Australia \\ ${}^2$ Department of
Electromagnetics, Chalmers University of Technology, S-41296
G\"oteborg, Sweden} \maketitle
\begin{abstract}
We apply the concepts of nonlinear guided-wave optics to a
Bose-Einstein condensate (BEC) trapped in an external potential.
As an example, we consider a parabolic double-well potential and
derive coupled-mode equations for the complex amplitudes of
the BEC macroscopic collective modes. Our equations describe
different regimes of the condensate dynamics, including the
nonlinear Josephson effect for any separation between the wells.
We demonstrate macroscopic self-trapping for both repulsive
and attractive interactions, and confirm our results by
numerical simulations.
\end{abstract}


\maketitle
\begin{multicols}{2}
\narrowtext 
A system of interacting bosons confined within an external
potential at zero temperature can be described by a macroscopic
wave function having the meaning of an order parameter and
satisfying the Gross-Pitaevskii (GP) equation \cite{review:1999}.
The GP equation is  a nonlinear equation that takes into account
the effects of the particle interactions through an effective mean
field, and it describes the condensate dynamics in a confined
geometry. Similar models of the confined dynamics of macroscopic
systems appear in other fields, e.g. in the case of an electron
gas confined in a quantum well, or optical modes of a photonic
microcavity \cite{Reith}. In all such systems, confined
single-particle states are restricted to discrete energies that
form a set of eigenmodes.

The physical picture of eigenmodes remains valid in the nonlinear
case \cite{Yuri:1999}, and {\em nonlinear collective modes}
correspond to the ground and  higher-order (excited) states of the
Bose-Einstein condensate (BEC) \cite{yukalov}. Moreover, it is
possible  to observe at least the first excited (antisymmetric)
collective mode experimentally \cite{exper}, through the collapses
and revivals in the dynamics of strongly  coupled two-component
BECs \cite{numerics}. The interest in the  non-ground-state
collective modes of BECs has grown dramatically with the study  of
 vortex states, very recently successfully created in the
experiment \cite{vortex}.

The modal structure of the condensate macroscopic (ground and excited)
 states  allows us to draw a deep analogy between BEC in a trap
and guided-wave optics, where the concept of nonlinear guided
modes is widely used \cite{supermodes}. The physical description of confined
condensate dynamics in time is akin to that of stationary beam
propagation along a nonlinear optical waveguide, with the BEC
chemical potential playing  the role of the beam propagation
constant. As is well known from nonlinear optics, the guided waves
become coupled in the presence of  nonlinearity, and the mode
coupling can lead to the nonlinear  phase shifting between the
modes, power exchange, and  self-trapping.

In this paper, {\em we apply the concepts of nonlinear
guided-wave optics to the  analysis of mode coupling and
intermodal population exchange in trapped BECs}. As the most
impressive (and also physically relevant) example of the
applications of our theory, we consider the BEC dynamics
 in a harmonic double-well potential, recently discussed in the
literature \cite{double}. We study the coupling between the
 BEC ground-state mode and the first excited (antisymmetric) mode 
in such a potential, and derive the dynamical equations for the
 the complex mode amplitudes, {\em valid for any value
of the well separation}. Our model comprises, in the limiting case
of large separation, the theory of Josephson tunneling developed for weakly interacting condensates in two separate
harmonic traps \cite{Smerzi}. In the limit of close separation,
our theory describes a  nonlinear population exchange between the
interacting modes, similar to the effective Rabi oscillations
in two-component BECs, studied both theoretically \cite{numerics} and
 experimentally \cite{exp1}.

We consider the macroscopic dynamics of BEC in
a strongly anisotropic external potential, $U=\frac{1}{2}m
\omega^2(Y^2+Z^2+\lambda X^2)$, created by a magnetic trap with a
characteristic frequency $\omega$. In the case of the cigar-shaped
trap, $\lambda \ll 1$, the collective BEC dynamics can be
described by a one-dimensional GP equation. Details of the
derivation and normalization can be found, e.g., in Refs.
\cite{perez}. The
 GP equation for the longitudinal profile of the normalized
condensate wave function  takes the form:
\begin{equation}
\label{GP1}
i\frac{\partial \psi}{\partial t} + \frac{\partial^{2}\psi}{\partial x^{2}}-
U(x) \psi + \sigma
|\psi|^{2} \psi = 0.
\end{equation}
According to normalization, the number of the condensate particles
${\cal N}$ is defined as  ${\cal N}=
(\hbar\omega/2U_{0}\sqrt{\lambda})N$, where $U_{0} = 4\pi
\hbar^{2}(a/m)$ characterizes two-particle interaction
proportional to the $s$-wave scattering length $a$, and the
functional $N = \int^{\infty}_{-\infty} |\psi|^{2}d x$ is the
integral of motion for the normalized nonstationary GP equation
(\ref{GP1}). The value of $\sigma =- {\rm sgn}(a) =\pm 1$ in front of
the nonlinear term is defined by the sign of the scattering length
of the two-body interaction, repulsive for $a>0$, and attractive
for $a<0$. The potential $U(x)=k(|x|-x_0)^2$ (below we take 
$k=1$) describes the double-well structure of the trap
in the longitudinal direction.

In the linear limit (i.e. for an ideal non-interacting gas), we
should consider $\sigma=0$, and the exact stationary solutions  of
Eq. (\ref{GP1}) in the form $\psi (x,t) =\Phi_j (x) e^{i\beta_j
t}$ are found in terms of the parabolic elliptic functions
\cite{merzbacher} that define a set of confined stationary states
existing at certain discrete values of $\beta_j$ ({\em linear
eigenmodes}). For $\sigma \neq 0$,  we can introduce, in a similar
way, a set of {\em nonlinear eigenmodes} \cite{Yuri:1999} described by real functions $\Phi_j(x)$ that satisfy the following nonlinear equation,
\begin{equation}
\label{statio}
\frac{d^2\Phi_j}{dx^2} + \beta_j \Phi_j - U(x)
\Phi_j + \sigma \Phi_j^3 =0.
\end{equation}
While in the case $\sigma=0$ the eigenvalue for each mode
$\beta_j$ is unique for any  given trap separation $x_0$, in the
nonlinear case there exist families of  localized solutions
$\Phi_{j}$ characterized by the dependence of the norm $N_j = \int
dx \Phi_j^2(x)$ on $\beta_j$, and the eigenvalue now becomes a
parameter of a {\em continuous family} \cite{Yuri:1999}. Figure
1 shows two examples of the ground-state mode $\Phi_0(x)$ and
first-order excited mode $\Phi_1(x)$ of the BEC with $N_{0}=N_{1}$
but $\beta_{0} \neq \beta_{1}$, for different values of $x_0$. The
dependencies $\beta_{0}$ and $\beta_{1}$ on the trap separation
$x_0$ are quite different for two signs of $\sigma$, as is seen in
Fig. 2.
\begin{figure}
\setlength{\epsfxsize}{8.0cm}
\centerline{\epsfbox{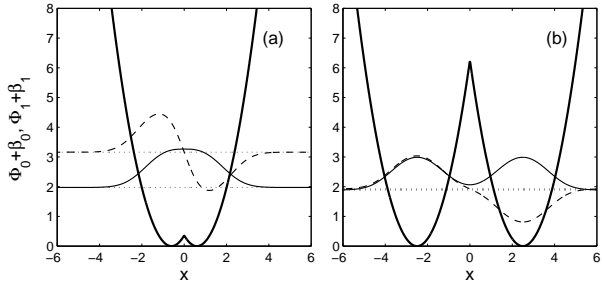}}
\caption{Confining potential (bold) with the ground (solid) and first
excited (dashed) collective modes for (a) $x_0=0.6$, and (b)
$x_0=2.5$
($\sigma =-1$, $N_0=N_1=5.0$). Dotted lines -- corresponding
values of the chemical potential: (a) $\beta_0 =1.974$, 
$\beta_1=3.162$,
 and (b) $\beta_0 =1.889$, $\beta_1=1.925$.}
\label{fig1}
\end{figure}
\begin{figure}
\setlength{\epsfxsize}{6.0cm}
\centerline{\epsfbox{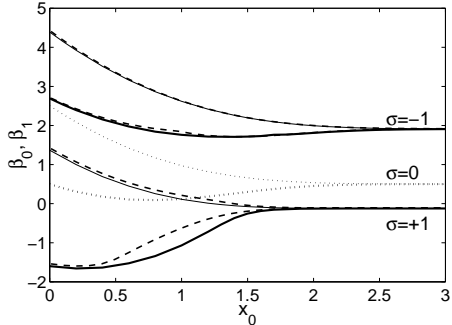}} \caption{Dependencies
$\beta_0(x_0)$ (lower curve) and $\beta_1(x_0)$ (upper curve) for $\sigma =0$ (dotted) and $\sigma =\pm 1$ (solid, $N_0=N_1=5.0$). Dashed - results of the
variational approach.} \label{fig2}
\end{figure}
To develop a coupled-mode theory for BECs, we consider the mode
interaction in a double-well potential and assume that the
condensate wave function is described by a superposition of two
modes of different symmetry, i.e. symmetric and antisymmetric,
\begin{equation}
\label{ansatz}
\psi(x,t) = b_0(t) \Phi_0(x) e^{-i\beta_0 t} + b_1(t) \Phi_1(x)
e^{-i\beta_1 t},
\end{equation}
where $b_j$  $(j=0,1)$ are the complex amplitudes, and $\Phi_j(x)$
may, in general, be any two solutions of Eq. (\ref{statio}). Then, the BEC
dynamics can be deduced from  the rate equations for the modal
amplitudes $b_{j}$. A similar approach has previously been
employed in Refs. \cite{Smerzi} to describe coherent tunneling
between two largely separated harmonic traps, with the basis
$\Phi_j$ consisting of the ground-state modes of individual potential wells. In contrary, our basis eigenfunctions are {\em the local nonlinear
modes of the entire double-well trap} which can be found exactly
for any given trap separation. Our approach is therefore similar
to the analysis of the power transfer between the cores of a
nonlinear optical coupler usually carried out in the terms of
the {\em supermodes}, i.e. the local modes of a composite core
\cite{supermodes}.

To derive the equations for the complex mode amplitudes $b_j(t)$,
we use the standard procedure, substituting the ansatz
(\ref{ansatz}) into the nonstationary GP equation (\ref{GP1}), and
averaging over the spatial dimension after multiplying the GP
equation by either $\Phi_0(x)$ or $\Phi_1(x)$. This yields a
system of two coupled equations,
\[
\begin{array}{l}
  {\displaystyle
  i \frac{d B_0}{dt} =  \sigma C_0 |B_0|^2 B_0 +  \sigma
  C_{01} (2|B_1|^2B_0 + B_0^*B_1^2
e^{-i\Omega t}),}
          \\*[9pt]
  {\displaystyle
i \frac{d B_1}{dt} =  \sigma C_1 |B_1|^2 B_1 + \sigma C_{01}
(2|B_0|^2B_1 + B_1^*B_0^2 e^{i\Omega t}).}
  \end{array}
\]
Here $\Omega =2(\beta_1 -\beta_0) - 2\sigma(C_0N_0 - C_1N_1)$,
and the coupling coefficients are defined as $C_j =
\gamma_{jj}/N_j^2$, $C_{01} =\gamma_{01}/(N_0 N_1)$, where
$\gamma_{ij} = \int dx \Phi_i^2(x) \Phi_j^2(x)$, and $B_j$ are the
normalized mode amplitudes, $B_j(t) = \sqrt{N_j} b_j(t) \exp
(-i\sigma C_{j}N_jt)$. These equations conserve the total norm
$|B_0|^2 + |B_1|^2 = n_0(t) + n_1(t) \equiv n$, where $n_0$ and
$n_1$ have the meaning of the time-dependent population numbers
for the two macroscopic states, and $n=N=N_0|b_0|^2+N_1|b_1|^2$.
It is important to note, that the form of the rate equations does
not depend on the normalization conditions for the basis
functions. For example, the condition
$\int^\infty_{-\infty}|\Phi_j|^2dx=\int^\infty_{-\infty}|\psi|^2dx=1$
 simply imposes the constraint $|B_j|=|b_j|$, so that
$n=N=|b_0|^2+|b_1|^2$.

Separating the amplitudes and phases as $B_j(t) =\sqrt{n_j(t)}
\exp [-i\phi_j(t)]$, we obtain a system of coupled equations for
the population difference of the two states $\Delta=n_1-n_0$, and
the relative phase shift, $\Theta =2(\phi_0-\phi_1) - \Omega t$,
\begin{equation}
\label{coupled2}
 \begin{array}{l}
  {\displaystyle
  \frac{d\Delta}{dt} = \sigma (n^2 -\Delta^2) \, \sin \Theta,}
          \\*[9pt]
  {\displaystyle
  \frac{d\Theta}{dt} = -\delta + \sigma(C_0+C_1)\Delta -
2\sigma C_{01}(2+ \cos \Theta)\Delta,}
  \end{array}
\end{equation}
where $\delta = 2(\beta_1
-\beta_0)+\sigma[(n-2N_0)C_0-(n-2N_1)C_1]$. System
(\ref{coupled2}) can be rewritten in a canonical form, $d\Delta/dt
=-\partial H/\partial \Theta$, $d\Theta/dt =
\partial H/\partial \Delta $,
with the Hamiltonian: $H=\sigma(n^2-\Delta^2)C_{01}\cos \Theta +
\sigma \left[ (C_0+C_1)/2 - 2C_{01} \right]\Delta^2 - \delta
\Delta.$ A mechanical analogy of this system may describe the motion
of a non-rigid pendulum with angular momentum $\Delta$ and a 
generalized angular coordinate $\Theta$. On the other hand, Eqs. (\ref{coupled2}) closely resemble the dynamic equations for the guided power and phase difference of two nonlinearly interacting orthogonally polarized optical modes in a birefringent fiber \cite{supermodes}. The exact solution of the
system (\ref{coupled2}) can be obtained in terms of elliptic
Jacobi functions and will be presented elsewhere.

The dynamics described by Eqs. (\ref{coupled2}) depends crucially on the values of the coupling coefficients $C_0$, $C_1$, and $C_{01}$, which are
determined by integration over the eigenmode profiles, so that
the results can be {\em quite different for the two signs of
$\sigma$}. Moreover, the condensate dynamics changes with the
separation of the potential wells, as governed by the dependencies
$C_0(x_0)$, $C_1(x_0)$, and $C_{01}(x_0)$, which differ drastically for $\sigma =\pm 1$ (see
Fig. 3). As expected from the linear theory \cite{merzbacher} and
 results for $\sigma=-1$ \cite{double}, the energy spectrum
becomes degenerate for large separation (see also Fig. 2), and the
coupling between the collective modes becomes more coherent.
Importantly, for $\sigma =+1$, this happens at the values of
separation smaller than those for $\sigma=-1$.
\begin{figure}
\setlength{\epsfxsize}{8.6cm} \centerline{\epsfbox{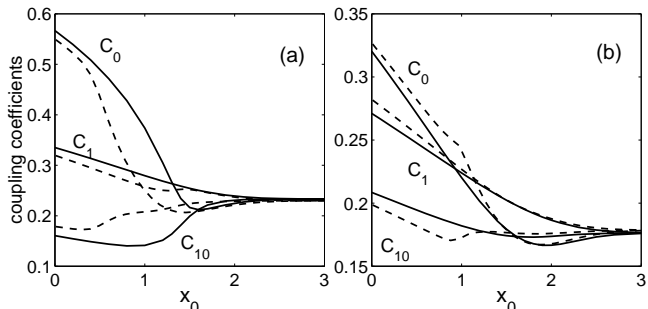}}
\caption{Coupling coefficients as functions of
separation for (a) $\sigma =+1$, and (b) $\sigma =-1$. Dashed -
results of the variational approach.} \label{fig3}
\end{figure}

The routine of calculating the coupling coefficients numerically
can be bypassed by employing {\em the variational approach}, using the
trial functions in the form of a linear superposition of the ground
states of isolated traps: $\Phi_{0,1}=A_{0,1}\{\exp
[(x-x_0)^2/2a^2_{0,1}] \pm \exp [(x+x_0)^2/2a^2_{0,1}]\}$, and the
Lagrangian of the stationary GP equation,
\[
{\mathcal{L}}=-\frac{1}{2}\left( \frac{d\Phi_j}{d x} \right)^2 +
\frac{1}{2}\left[ \beta_j - U(x)\right] \Phi^2_j -
\frac{1}{4}\sigma\Phi^4_j.
\]
By inserting the trial functions into the corresponding variational
integral, an explicit, although algebraically complicated
integrated Lagrangian is obtained. The variational
equations with respect to the parameters $A_{0,1}$ and $a_{0,1}$
yield relations which determine the characteristic eigenvalues
$\beta_{0,1}(x_0)$ and the coefficients $C_0$, $C_1$, and
$C_{01}$. These relations can be further simplified in the limit
$x_0 \gg 1$, but must be solved numerically for a general case.
Comparisons between the variational predictions and the results obtained by solving Eq. (\ref{statio}) numerically are shown in Fig. 2, for
$\beta_{0,1}(x_0)$, and in Fig. 3, for  $C_0$,
$C_1$, and $C_{01}$. The agreement is seen to be satisfactory.

To visualize the population dynamics, in Fig. 4 we plot the phase
portraits $\{\Theta,\Delta\}$ of the dynamical system (\ref{coupled2}) for the case
$\sigma=+1$ and $N_0=N_1$. For convenience, the population
difference, $\Delta$, is measured in the units of $n$. For small
separations [Figs. 4(a,b)], while the coupling coefficients are
sufficiently different, there are only two fixed states of the
relative population: $\Delta = \pm n$, which corresponds to either
$n_0=0$ or $n_1=0$. In both these states, the phase is unbounded,
i.e. it is a linear function of time. The other phase trajectories in Figs. 4(a,b) represent the dynamical states with the {\em running phase}, that is a 
delocalized phase. The mechanical analogy of this phenomenon is simple \cite{Smerzi}: it corresponds to a self-sustained steady closed-loop rotation of a 
non-rigid pendulum around its support. In terms of the condensate dynamics, 
these states describe the nonlinear Rabi-type oscillations 
between the ground and first excited macroscopic states, for small $x_0$, and 
the Josephson-type tunneling between the two potential wells, for a
 sufficiently large separation. Remarkably, the population of either 
well is never completely depleted.

\begin{figure}
\setlength{\epsfxsize}{8.5cm} \centerline{\epsfbox{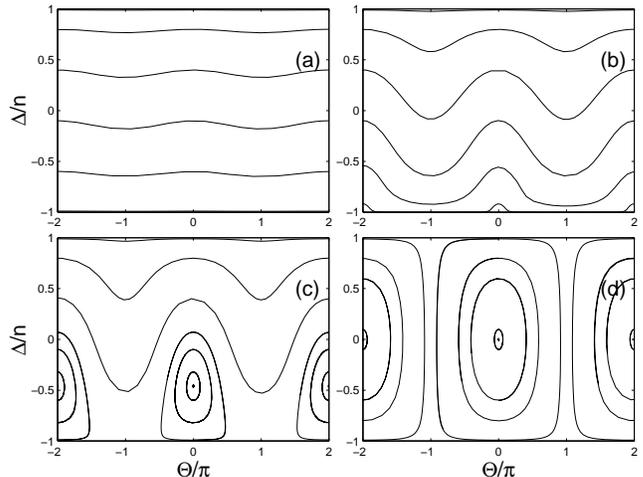}}
\caption{(a)-(d) Phase plane $\Delta(\Theta)$ for Eqs.
(\ref{coupled2}) at $\sigma =+1$, $N=n=1$, and $x_0=0.0, 1.5,
1.7$, and 
$3.0$, respectively.} \label{fig4}
\end{figure}
The phase plane shown in Fig. 4 also reveals the existence of macroscopic quantum self-trapped
(MQST) states, predicted and described in \cite{Smerzi} for 
{\em weakly interacting} BECs in largely separated 
traps. The MQST states are characterized by a nonzero average
population imbalance. As the separation between the wells grows, a bifurcation of the fixed points on the phase plane $\{\Delta, \Theta \}$ occurs, and the
stable centers, corresponding to the MQST states with a {\em
trapped phase}, appear [see Figs. 4(c,d)]. This occurs at a certain $x^{cr}_0$, for which the condition $\delta=n(6C_{01}-C_0-C_1)$ is satisfied. For $\sigma=+1$ and $n=1$, for example, $x^{cr}_0 \simeq 1.48$ which agrees very well with the corresponding result of the variational approach, $x^{cr}_0 \simeq 
1.41$. For larger separation,
when $C_0 \sim C_1 \sim C_{01}$, the positions of the centers are
approximately given by: $\Delta \approx
(\beta_0-\beta_1)/(2 C_{01} \sigma)$ at $\Theta=\pm(2m)\pi$, 
where $m$ is integer.
With increasing separation, as $(\beta_0-\beta_1) \to 0$, these
fixed centers move towards the line $\Delta=0$, and the saddles
form between them, so that the MQST states, other than those
identical to the ground states of the individual wells, cease to
exist [see Fig. 4(d)]. The oscillations of the population
imbalance around the stable fixed points with the trapped phase
have been identified in \cite{Smerzi} as  $\pi$-states. Clearly,
the effect we observe here is {\em qualitatively similar}, except
for the average value of the trapped phase which, due to a
different choice of the basis eigenfunctions and definition of $\Theta$, is equal to $\pm(2m)\pi$.
\begin{figure}
\setlength{\epsfxsize}{7.0cm}
\centerline{\epsfbox{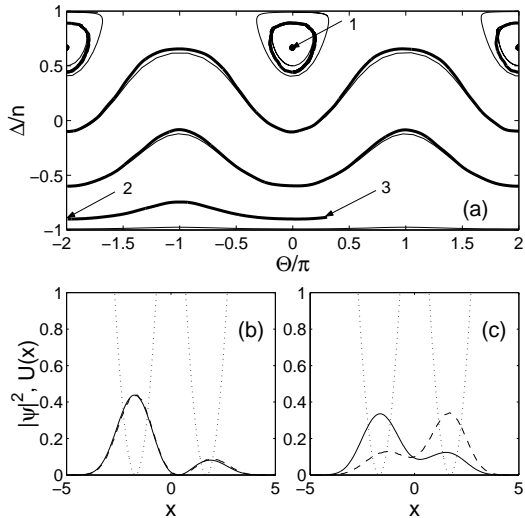}}
\caption{(a) Phase trajectories $\Delta(\Theta)$ calculated from
Eqs. (\ref{coupled2}) for $\sigma=-1$, $x_0=1.7$ (thin) compared
with the numerical solution of the GP equation (thick). (b)
$|\psi|^2$ in the MQST state $''1''$ in (a), at the normalized
times $t=0$ (solid) and $t=100$ (dashed). (c) $|\psi|^2$
corresponding to the point $''2''$ in (a) at $t=0$ (solid), and
corresponding to the point $''3''$ in (a) at $t=20$ (dashed).
Double-well potential is shown in (b,c) by a dotted curve.} \label{fig5}
\end{figure}

To compare the predictions of our coupled-mode theory with the
actual dynamics of the BEC in a double-well potential modeled by
the GP equation, we solve Eq. (\ref{GP1}) numerically. As an
initial condition, for both $\sigma = \pm 1$, we take
$\psi(x,0)=b_0(0)\Phi_0(x)+b_1(0)\Phi_1(x)$, with
$b^2_0(0)+b^2_1(0)=1/{N_0}=1/{N_1}$. In Fig. 5(a), the phase
trajectories $\Delta(\Theta)$ are compared with those calculated
using Eqs. (\ref{coupled2}), for $\sigma=-1$, and the trap separation corresponding to sufficiently dissimilar values of $\beta_j$ and $C_j$ (see Figs. 2 and 3). It is clear that the approximate equations of the coupled-mode theory
correctly describe the dynamics of the condensate in the
states with a running phase [see Fig. 5(c)], as well as the
position of the MQST states, one of which is shown in Fig. 5(b).
Performing a similar comparison for different $\sigma$ and $x_0$, we can conclude that the eigenfunctions $\Phi_j(x)$ represent a good basis
for the modal decomposition of the macroscopic condensate
wave function $\psi(x,t)$. The adiabatic evolution of the
eigenfunctions with time, although leading to slight deviations of
the condensate states from the exact MQSTs [see Fig. 5(a)], does
not introduce significant damping into the system, and therefore
does not lead to a dramatic switching between the states.

In conclusion, we have employed the concepts of the nonlinear
guided-wave optics and  developed, for the first time to our
knowledge, a consistent coupled-mode theory for BECs. We have
studied the BEC dynamics in a double-well harmonic trap, and
verified the results by numerical simulations of the
nonstationary GP equation. The strong advantage of our theory is
its ability to describe the condensate dynamics for {\em any well
separation}, including the Josephson tunneling effect at large
separations, mode coupling and Rabi oscillations in a single harmonic
well, and the macroscopic self-trapped states in the crossover
regime.

\end{multicols}


\vspace{-3mm}
\begin{references}
\bibitem{review:1999} F. Dalfovo, S. Giorgini, L. P. Pitaevskii, and S.
Stringari, Rev. Mod. Phys. {\bf 71}, 463 (1999).

\bibitem{Reith} J. P. Reithmaier, M. R{\"o}hner, H. Zull, F. 
Sch{\"a}fer, A. Forchel, P. A. Knipp, and T. L. Reinecke, Phys. Rev. Lett. {\bf 78}, 378 (1997); M. Bayer, T. Gutbrod, J. P. Reithmaier, A. Forchel, T. L. 
Reinecke, P. A. Knipp, A. A. Dremin, and V. D. Kulakovskii, Phys. Rev. Lett. {\bf 81}, 2582 (1998).

\bibitem{Yuri:1999} See, e.g., Yu.S. Kivshar, T.J. Alexander, and S.K.
Turitsyn, cond-mat/9907475.

\bibitem{yukalov} V. I. Yukalov, E. P. Yukalova, and V. S. Bagnato,
 Phys. Rev. A {\bf 56}, 4845 (1997).

\bibitem{exper} J. Williams, R. Walser, J. Cooper, E. A. Cornell, and  
M. Holland, cond-mat/9904399.

\bibitem{numerics} J. Williams, R. Walser, J. Cooper, E. Cornell, and M.
Holland, Phys. Rev. A {\bf 59}, R31 (1999).

\bibitem{vortex} M. R. Matthews, B. P. Anderson, P. C. Haljan, D. S. Hall, 
C. E.
Wieman, and E. A. Cornell, Phys. Rev. Lett. {\bf 83}, 2498 (1999).

\bibitem{supermodes} See, e.g., A. Vitarescu, Appl. Phys. Lett. {\bf 
49},
61 (1986); Y.  Silberberg and G. I. Stegeman,  Appl. Phys. Lett.
{\bf 50}, 801 (1987).

\bibitem{double} P. Capuzzi and E. S. Hern{\'a}ndez, Phys. Rev. A {\bf 
59},
1488 (1999); cond-mat/9902140; R. W. Spekkens and J. E. Sipe, Phys. Rev. 
A
{\bf
59}, 3868 (1999); N. Tsukada, M. Gotoda, Y. Nomura, and T. Isu, Phys. 
Rev. A {\bf 59}, 3862 (1999).

\bibitem{Smerzi} A. Smerzi, S. Fantoni, S. Giovanazzi, and S. R. Shenoy, 
Phys.
Rev. Lett. {\bf 79}, 4950 (1997); S. Raghavan, A. Smerzi, S. Fantoni, 
and S.R.
Shenoy, Phys. Rev. A {\bf 59}, 620 (1999); I. Marino, S. Raghavan, S. 
Fantoni,
S.R. Shenoy, and A. Smerzi,  Phys. Rev. A {\bf 60}, 487 (1999).

\bibitem{exp1} M. R. Matthews, B. P. Anderson, P. C. Haljan, D. S. Hall, 
M. J.
Holland, J. E Williams, C. E. Wieman, and E. A. Cornell, cond-mat/9906288.

\bibitem{perez} V. M. Perez-Garcia, H. Michinel, and H. Herrero, Phys. 
Rev. A
{\bf 57}, 3837 (1998); Yu.S. Kivshar and T. J. Alexander, 
cond-mat/9905048.

\bibitem{merzbacher} E. Merzbacher, {\em Quantum Mechanics} (Wiley, New 
York,
1961), pp. 65-78.




\end{references}
\end{document}